\documentclass[aps,showpacs]{revtex4}
\usepackage{graphicx}
\def\bc{\begin{center}}
\def\ec{\end{center}}

\def\beq{\begin{equation}}
\def\eeq{\end{equation}}

\def\bk{{\bf k}}

\def\br{{\bf r}}

\begin{document}

%e-mail: ziegler@physik.uni-augsburg.de\\
%telephone: (49) 821 598 3244, FAX: (49) 821 598 3262

%\maketitle
%Title of paper
\title{Metal-insulator transition in three-dimensional semiconductors\\
%{\small (3d mit.tex)}
}

\author{K. Ziegler}
%\affiliation{Institut f\"ur Physik, Universit\"at Augsburg\\
%D-86135 Augsburg, Germany}
\date{\today}

\begin{abstract}
We use a random gap model to describe a metal-insulator transition in three-dimensional semiconductors
due to doping and find a conventional phase transition, where the effective scattering rate is the order
parameter. Spontaneous symmetry breaking results in metallic behavior, whereas the insulating
regime is characterized by the absence of spontaneous symmetry breaking. The transition is continuous
for the average conductivity with critical exponent equal to 1. Away from the critical point the 
exponent is roughly 0.6, which may explain experimental observations of a crossover of the exponent 
from 1 to 0.5 by going away from the critical point. 
\end{abstract}

\maketitle

\section{Introduction}

Undoped semiconductors have a small gap between the valence and the conduction band, typically
of the order of $0.2 ... 1.2$ eV \cite{mermin76}. This gap is strongly affected by doping.
In particular, sufficiently strong doping closes the gap such that a metallic phase appears.
A classical example for this type of metal-insulator transition is doped silicon, where typical 
dopants are phosphorus (Si:P) or boron (Si:B) \cite{rosenbaum80,rosenbaum84,roy88,rosenbaum94,loehneysen98}.  
Disorder plays a crucial role in these materials due to the inhomogeneous distribution of the dopants. 
This suggested that Anderson localization must play a crucial role in these systems, where the
quantum states would undergo a transition from extended to localized states for increasing disorder. 
This transition should be reflected in the transport properties, where extended states lead to a metal
and localized states to an insulator at vanishing temperatures.

Measurements of the conductivity $\sigma(N)$ as a function of doping concentration
$N$ in Si:P at low temperatures has indeed revealed a critical behavior. Above a critical concentration 
$N_c$ a power law was found
\[
\sigma(N)\sim \sigma_0(N/N_c-1)^\mu \  \  \ (N\ge N_c)
\]
and a vanishing conductivity for $N<N_c$. The exponent $\mu$ was determined as $\mu\approx0.5$ for some
experiments \cite{rosenbaum80,rosenbaum84,roy88,rosenbaum94}, whereas a crossover from $\mu\approx0.5$
at some distance from the critical point to $\mu\approx1$ in a vicinity very close to $N_c$ was observed in
other experiments \cite{rosenbaum94,loehneysen98}. 

Although the picture of an Anderson transition is quite appealing, an alternative description can be provided
by a random gap model. The idea is that the dopants create energy levels inside the semiconductor gap.
These levels are associated with states that can overlap with the states in the semiconductor bands and 
eventually fill the semiconductor gap by forming extended states. The effect can be described by a random distribution 
of local gaps. Then the locally filled gaps can be distributed over the entire system  and form eventually, after sufficient
doping, a conducting ``network''. This  is associated with a second-order phase transition which will be described in 
this articles. The transition is distinguished
from the Anderson transition by the fact that the metallic phase appears at strong disorder (i.e. high dopant 
concentration) and the insulating phase at weak disorder. This does not rule out an Anderson transition if we
increase the disorder inside the metallic regime. However, in realistic systems it is more likely to see the
transition caused by the random gap than the more sophisticated Anderson transition.

\section{Model and symmetries}

We consider a two-band model with a symmetric Hamiltonian. This can be expressed in terms of Pauli matrices $\sigma_j$
($j=0,...,3$). % (i.e. not Dirac!):
A simple case is
\beq
H=h_1\sigma_1+h_3\sigma_3 %, \ \ \ (e.g. \ \ h_1=k)
\label{hamiltonian00a}
\eeq
with symmetric matrices $h_1$, $h_3$ in three-dimensional (real) space. To be more specific, we can 
choose the Fourier components $h_1=k/\sqrt{2m}$ with
$k \equiv\sqrt{k_1^2+k_2^2+k_3^2}$. For a uniform gap $h_3=\Delta$ we obtain the dispersion 
$E_\bk=\pm\sqrt{k^2/2m+\Delta^2}$. For small wavevector $k$ we can approximate the spectrum as 
$E_\bk\sim\pm [\Delta+k^2/2m\Delta +o(k^4)]$, where the gap $\Delta$ renormalizes the
quasiparticle mass $m$. Subsequently we will consider a random gap $h_3$ to describe
the effect of an inhomogeneous distribution of dopants.  

The one-particle Hamiltonian $H$ is invariant under an Abelian chiral transformation:
\beq
e^{\alpha\sigma_2}He^{\alpha\sigma_2}=H
\ .
\label{abelian0}
\eeq
In order to reveal the relevant symmetry for transport in this system, we construct the two-body Hamiltonian
\[
{\hat H}=\pmatrix{
H & 0 \cr
0 & H \cr
},
\]
where the upper block $H$ acts on bosons, the lower block $H$ on fermions.
The reason for introducing this two-body Hamiltonian is that we can transform
the distribution of the random Hamiltonian $H$ into a distribution of the Green's
function ${\hat G}(z)=({\hat H}-z)^{-1}$ \cite{ziegler97,ziegler09a}, which is
often called a supersymmetric representation of the Green's function \cite{efetov}.

Next we introduce the transformation matrix
\beq
{\hat U}=\pmatrix{
0 & \varphi\sigma_2 \cr
\varphi'\sigma_2 & 0 \cr
}
\label{chiral1a}
\eeq
and obtain the anti-commutator relation
\beq
[{\hat H},{\hat U}]_+=0
\ .
\label{relations3a}
\eeq
This implies the non-Abelian chiral symmetry
\beq
e^{\hat U}{\hat H}e^{\hat U}={\hat H}
\ ,
\label{na_chirala}
\eeq
which is an extension of the Abelian symmetry (\ref{abelian0}). The Green's function ${\hat G}(z)$ does not
obey this symmetry for $z\ne0$. Therefore, $z$ plays here the role of a symmetry-breaking field. An interesting
limit is $z\to0$, which we will study in the next section.

Now we consider the case of a random gap with mean $\langle h_{3,\br}\rangle=\Delta$ and variance 
$\langle h_{3,\br}h_{3,\br'}\rangle-\Delta^2=g\delta_{\br,\br'}$ and its
effect on the average conductivity at frequency $\omega$. The conductivity is obtained from the Kubo
formula as \cite{altshuler,ziegler09a}
\beq
\sigma_{kk}=-\frac{e^2}{2h}\omega^2\lim_{\epsilon\to0}
Re\left\{\sum_\br r_k^2Tr_2\left[\langle G_{0\br}(\omega/2+i\epsilon) G_{\br0}(-\omega/2-i\epsilon)\rangle\right]
\right\} ,
\ \ \ G(z)=(H-z)^{-1}
\ .
\label{cond00}
\eeq
In particular, we are interested in the DC limit $\omega\to0$. This limit restores the chiral symmetry 
of ${\hat H}$ in (\ref{na_chirala}) for the Green's functions. However, the symmetry can be spontaneously broken now. Since it 
is a continuous symmetry, this creates a massless mode, which represents fluctuations on arbitrarily 
large length scales. 

Here it should be noticed that $\sigma_2(H+z)^{-1}\sigma_2=-(H-z)^{-1}$.
This has the consequence that the product in (\ref{cond00}) reads 
$G_{0\br}(z)G_{\br0}(-z)=(H- z)^{-1}_{0\br}(H+z)^{-1}_{\br0}=-(H-z)^{-1}_{0\br}\sigma_2(H-z)^{-1}_{\br0}\sigma_2$
such that elements of ${\hat G}(\omega/2+i\epsilon)$ are sufficient to express the conductivity.

\section{Self-consistent approximation}
\label{sect:self-consist}

\begin{figure}
\begin{center}
\includegraphics[width=8cm,height=7cm]{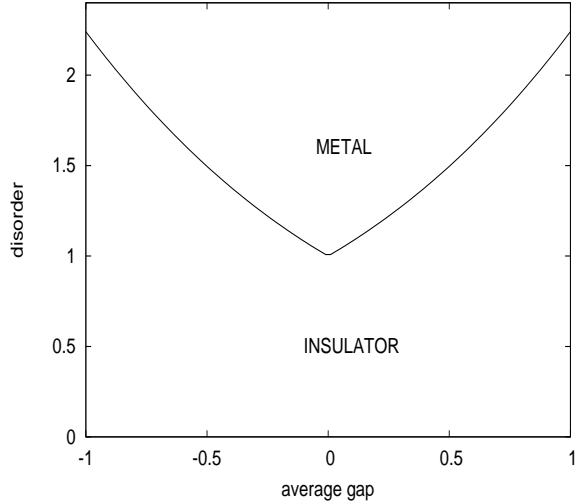}
\caption{
Phase diagram of the metal-insulator transition of the three-dimensional random gap model from Eq. (\ref{pd}),
where disorder is the parameter $\gamma$ and the average gap is $\Delta$. 
%The insulating - $\eta=0$ - (metallic - $\eta\ne0$ -) regime is below (above) the phase boundary.
}
\label{fig:phase_d}
\end{center}
\end{figure}

\begin{figure}
\begin{center}
\includegraphics[width=8cm,height=7cm]{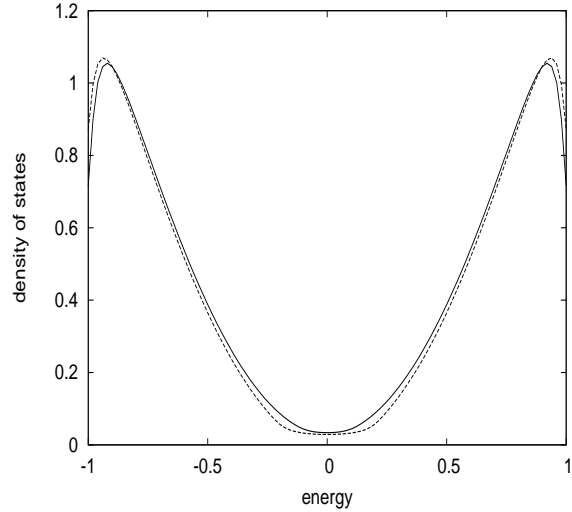}
\caption{
Average density of states of the three-dimensional random gap model for fixed $\eta=0.04$ and
average gap $\Delta=0.4$ (full curve) and $\Delta=0.8$ (dashed curve).}
\label{fig:dos}
\end{center}
\end{figure}

The self-consistent Born approximation of the average one-particle Green's function reads
\beq
\langle G(z)\rangle\approx G_0(z+i\eta) , \ \ \ G_0(z)=(\langle H\rangle -z)^{-1}
\ ,
\label{scba}
\eeq
where the self-energy $\eta$ is a scattering rate, which is determined by the self-consistent equation
$i\eta= G_{0,0}(z+i\eta)$ \cite{mahan}. This reads in our case
\[
i\eta=\gamma(z+i\eta)\left[\lambda-\frac{\alpha }{2}
\log\left(\frac{\alpha +\lambda}{\alpha -\lambda}\right)\right] 
\ \ \ (\gamma=g/2\pi^2,\ \ \alpha =\sqrt{(z+i\eta)^2-\Delta^2/4})
\]
and for $z=0$ this simplifies to $\eta=\eta I$ with
\[
I=\gamma\left[
\lambda-\beta\arctan(\lambda/\beta)\right],\ \ \ \beta=\sqrt{\eta^2+\Delta^2/4} 
\ .
\]
In this case
%the limit $z\to0$ 
there are two solutions of the self-consistent equation, namely $\eta=0$ and $\eta\ne0$ with
\beq
\gamma=\frac{1}{\lambda-\beta\arctan(\lambda/\beta)}
\ .
\label{spe0}
\eeq
A nonzero $\eta$ reflects spontaneous symmetry breaking with respect to
(\ref{na_chirala}). Such a solution exists for (\ref{spe0}) only at
sufficiently large $\gamma$. Moreover, $\eta$ vanishes continuously as we reduce $\gamma$. Then there 
is a phase boundary which separates the symmetric and the symmetry-broken regime:
\beq
\gamma(\Delta)=\frac{2}{2\lambda-\Delta\arctan(2\lambda/\Delta)}
\label{pd}
\eeq
which is plotted in Fig. \ref{fig:phase_d}.
The average density of states then reads
\[
\rho(E)=\frac{1}{2\pi}\lim_{\epsilon\to0}Im\left\{Tr_2\left[\langle G_{\br\br}(E+i\epsilon)\rangle\right]\right\}
\approx \frac{1}{2\pi}\lim_{\epsilon\to0}Im\left\{Tr_2 G_{0,0}(E+i\epsilon+i\eta)\right\}
\]
\beq
=\frac{1}{\pi}Im\left\{(E+i\eta)\left[\lambda-\frac{\alpha }{2}
\log\left(\frac{\alpha +\lambda}{\alpha -\lambda}\right)\right]\right\} ,
\ \ \ \alpha =\sqrt{(E+i\eta)^2-\Delta^2/4})
\ .
\label{dos1}
\eeq
As a qualitative picture the average density of states is plotted for a fixed $\eta$ in Fig. \ref{fig:dos}.

\section{Scaling relation of the average two-particle Green's function}

A common approximation for the conductivity in (\ref{cond00}) is the factorization 
of the averaged product of Green's functions as  \cite{altshuler}
\beq
\sum_\br r_k^2Tr_2\left[\langle G_{0\br}(y) G_{\br0}(-y)\rangle\right]
\approx\sum_\br r_k^2Tr_2\left[\langle G_{0\br}(y)\rangle\langle G_{\br0}(-y)\rangle\right]
\ \ \ (y=\omega/2+i\epsilon)
\ ,
\label{factor0}
\eeq
which can be combined with the self-consistent Born approximation in Eq. (\ref{scba}) to obtain
\beq
\sum_\br r_k^2Tr_2\left[\langle G_{0\br}(y)\rangle \langle G_{\br0}(-y)\rangle\right]
\approx \sum_\br r_k^2Tr_2\left[G_{0,\br}(y+i\eta)G_{0,-\br}(-y-i\eta)\right]
\ .
\label{drude0}
\eeq
For the expression (\ref{cond00}) this approximation leads to the Boltzmann (or Drude) conductivity, which reads in 
our specific case 
\beq
\sigma_{kk}
%\approx -\frac{e^2}{2h}\omega^2\sum_rr_k^2Tr_2\left[G_{0,r}(y+i\eta)G_{0,-r}(-y-i\eta)\right]
\approx \frac{e^2}{2h}\frac{\omega^2}{\pi^2}\int_0^\lambda  
\frac{\Delta^2/4 -z^2}{(\Delta^2/4-z^2+k^2)^3}k^2dk \ \ \ (z=\omega/2+i\eta)
\ .
\eeq
Thus the conductivity vanishes in the DC limit $\omega\to0$ for $\eta\ge0$.
% because of the integral
%\[
%\int_0^\lambda \frac{-y^2}{(-y^2+k^2)^3}k^2dk
%=-\frac{1}{16y}\log\left(\frac{k+y}{k-y}\right)\Big|_0^\lambda
%\ .
%\]
The reason is that the self-consistent Born approximation creates the Green's function 
$G_{0,\br}(y+i\eta)$, which decays exponentially on the scale $1/\eta$. Consequently, the 
sum over the real space on the right-hand side of Eq. (\ref{drude0}) is finite.

A more careful inspection indicates that the averaged product of Green's function on the left-hand side
of Eq. (\ref{factor0}) decays according to a power law as a consequence of the massless 
fluctuations around the spontaneous symmetry breaking solution $\eta\ne 0$ \cite{ziegler97}.
We can perform the integration with respect to these fluctuations and obtain the relation \cite{ziegler09a}
\beq
\sum_\br r_k^2Tr_2\left[\langle G_{0\br}(y) G_{\br0}(-y)\rangle\right]
%=f(y)\sum_rr_k^2 Tr_2\left[G_{0,r0}(iz(y))G_{0,0r}(iz(-y))\right]
=f(\eta/y)\sum_\br r_k^2 Tr_2\left[G_{0,r}(y+i\eta)G_{0,-r}(-y-i\eta)\right]
\ ,
\label{scaling_gf2}
\eeq
where the coefficient depends on the ratio of the order 
parameter of spontaneous symmetry breaking $\eta$ and the symmetry-breaking field $y$:
\beq
f(\eta/y) %=\frac{(y+i\eta)^2}{y^2}
=(1+i\eta/y)^2 %, \ \ \ {\bar\eta}=[\eta(\mu)+\eta(-\mu)]/2
\ .
\eeq
This coefficients represents correlations of the Green's function fluctuations, which are negligible only
for $f(\eta/y)\approx 1$. In the absence of symmetry breaking is $\eta=0$ and $f(0)=1$. This justifies
the approximation by Eq. (\ref{drude0}) in the insulating regime. 
% as it should be according to the relation (\ref{scaling_gf2}). 
Moreover, $f(\eta/y)$ diverges in the presence of spontaneous symmetry breaking for a vanishing 
symmetry-breaking field $y$. In particular, it is finite for $\omega>0$ which reflects the fact that
fluctuations are cut-off on length scales $L_\omega\sim v_F/\omega$ ($v_F$ is the Fermi velocity).
 
\begin{figure}
\begin{center}
\includegraphics[width=8cm,height=7cm]{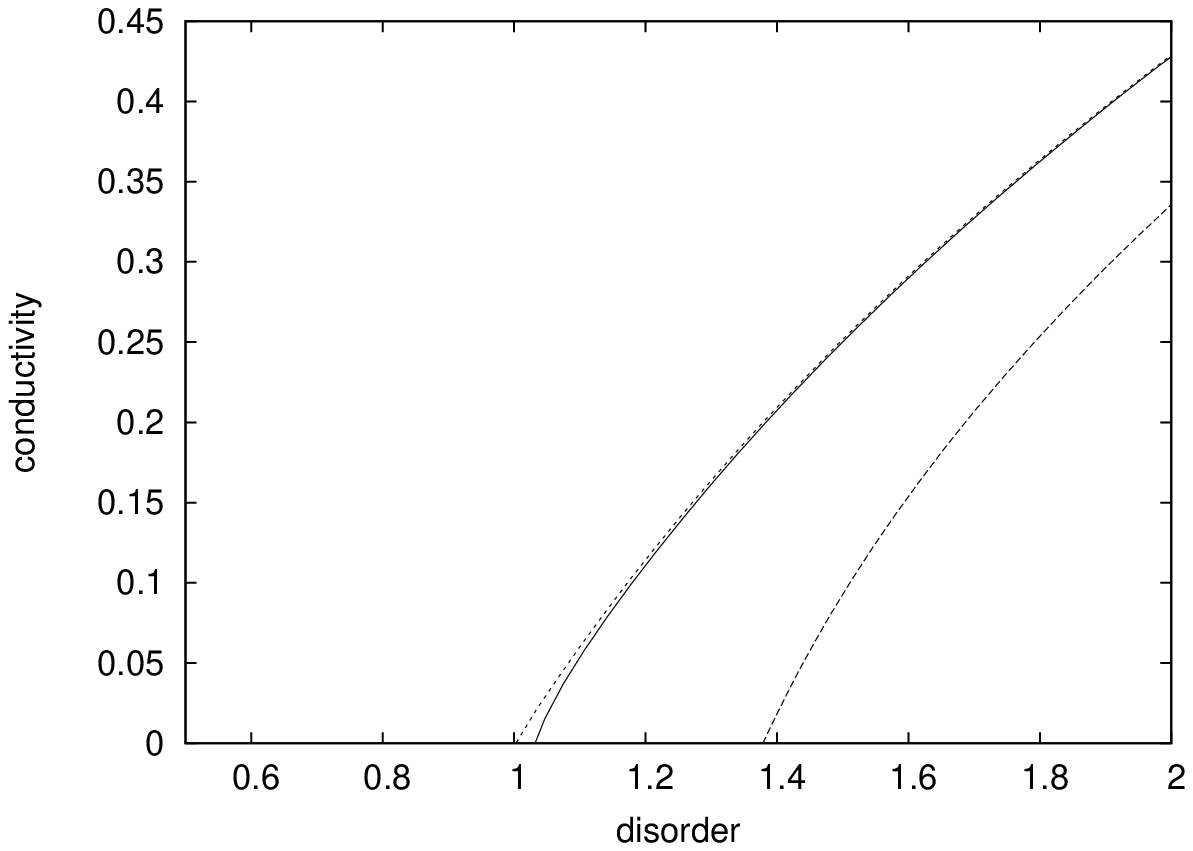}
\includegraphics[width=8cm,height=7cm]{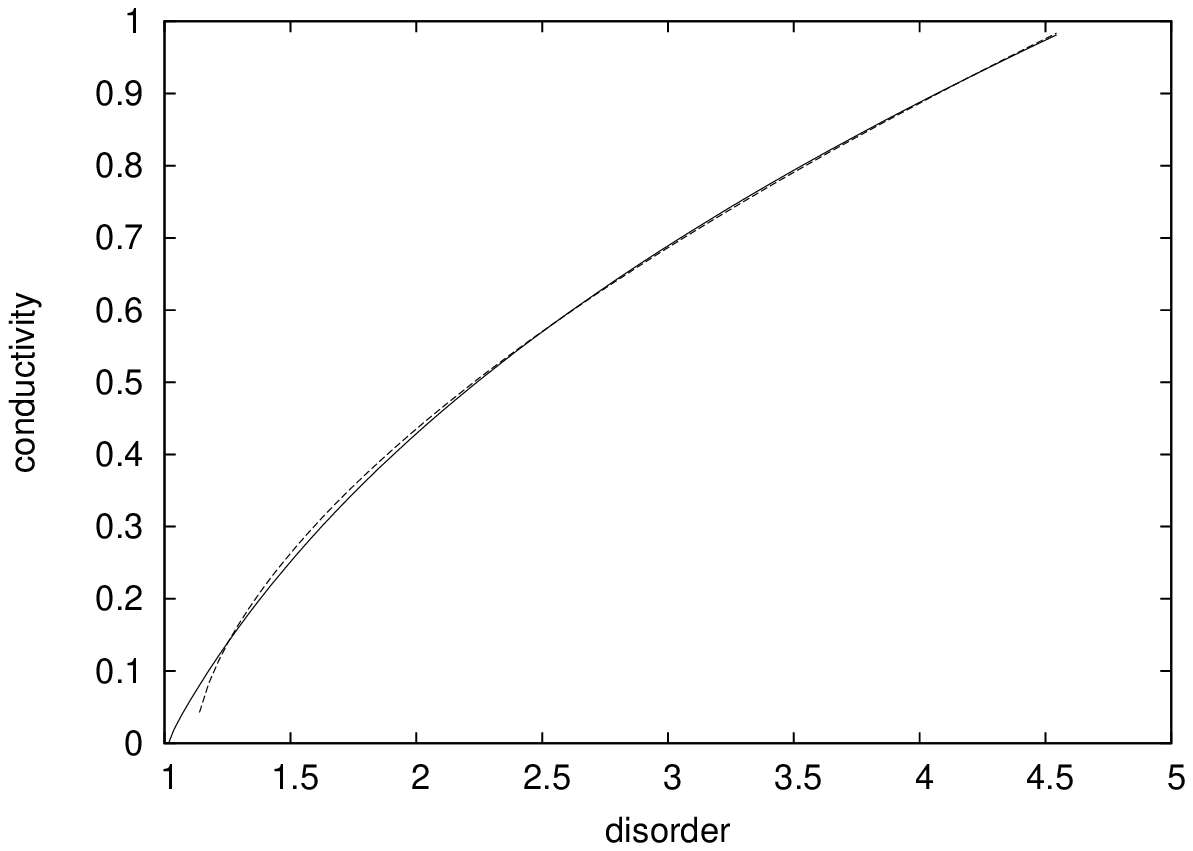}
\caption{
Conductivity as a function of disorder for an average gap $\Delta=0.004$ (dotted curve), $\Delta=0.04$ (full curve) 
and  $\Delta=0.4$ (dashed curve).
There is a metal-insulator transition at $\gamma\approx1$, at $\gamma\approx1.03$  and at $\gamma\approx 1.37$, respectively. %cond13a.mws
The second plot demonstrates the fit (dashed curve) of the conductivity (full curve) away from the critical point $\gamma_c$ 
for $\Delta=0.01$ by $0.47(\gamma-\gamma_c)^{0.6}$.
}
\label{fig:cond}
\end{center}
\end{figure}

With the scaling relation (\ref{scaling_gf2}) the conductivity in Eq. (\ref{cond00}) becomes 
in the DC limit $\omega\to0$
\beq
\sigma_{kk}
%=\frac{e^2}{2h}\eta^2\lim_{\epsilon\to0}Re\left\{\sum_\br r_k^2 Tr_2\left[G_{0,\br}(i\eta+y)G_{0,-\br}(-i\eta-y)\right]\right\} 
=\frac{e^2}{4\pi h}\frac{\eta^2}{\sqrt{\Delta^2/4+\eta^2}}
\ .
\label{cond1}
\eeq
The solution $\eta$ of the self-consistent Eq. (\ref{spe0}) is inserted into $\sigma_{kk}$
and the conductivity is plotted as a function of disorder strength $\gamma$ in Fig. \ref{fig:cond}. The conductivity vanishes
linearly with decreasing disorder strength (i.e. with decreasing doping concentration).

\section{Discussion and conclusions}

Our result for the DC conductivity in Eq. (\ref{cond1}), together with the solution of the order parameter $\eta$
in Eq. (\ref{spe0}), provides a simple description of a metal-insulator transition in doped three-dimensional 
semiconductors. The metal-insulator transition is characterized by the scattering rate $\eta$ that vanishes in the 
insulating regime. Such a behavior is not an Anderson transition, since the latter would have a scattering rate 
$\eta\ne0$ on both sides of the transition \cite{schaefer80}.
Even more important is the change of the coefficient $f(\eta/y)$: It is always 1 in the insulating regime
and infinite in the metallic regime. This quantity describes the correlations of the Green's function fluctuations in
the relation (\ref{scaling_gf2}).

There is a linear behavior near the metal-insulator transition and a crossover to a non-critical power law, % with an exponent $\mu\approx 0.6$,
as depicted in Fig. \ref{fig:cond}. 
For the linear part the slope of the conductivity is quite robust with respect to the average gap $\Delta$ (cf. first plot in Fig. 
\ref{fig:cond}). %This could be an advantage for applications of semiconductors as electronic switching and amplification devices. 
Away from the transition point a negative curvature appears though, which can be fitted by a power law with exponent $\mu\approx 0.6$ (cf. second 
plot in Fig. \ref{fig:cond}). 
The change of exponents can be related to the discussion in Refs. \cite{stupp93,loehneysen98} about 
a crossover of exponents in Si:P from $\mu\approx 1$ very close to the critical point $N_c$ to $\mu\approx0.5$ further away from $N_c$. 
Rosenbaum et al. have found that the conductivity close to the critical point varies from sample to sample \cite{rosenbaum94}. This indicates strong 
conductivity fluctuations, which may also exist in our random gap model, as indicated by the strong fluctuations of the Green's 
functions due to the large values of $f(\eta/y)$. % in Eq. (\ref{scaling_gf2}) are a signature for correlated fluctuations. 

The conductivity always increases with $\eta$. This may be an artifact of our approximation because we have neglected 
(massive) symmetry-breaking modes in the derivation of the scaling relation Eq. (\ref{scaling_gf2}). Symmetry-breaking modes can
become unstable for sufficiently strong disorder $\gamma>\gamma_i>\gamma_c$. 
These unstable modes terminate the validity of 
our approach, as we have seen for two-dimensional Dirac fermions with random gap \cite{ziegler97}. Such an instability might be linked
to Anderson localization. If this is true, it would be of interest to increase disorder in the metallic phase of doped semiconductors 
to observe a genuine Anderson transition. Another interesting aspect in these materials is the insulating phase, where magnetic correlations 
play a crucial role \cite{bhatt82,kettemann09}. This would require an extension of the random gap model by including spin degrees of freedom.

\end{document}